\documentclass[sigconf,10pt]{acmart}

\usepackage{url}
\usepackage{xspace}
\usepackage{etoolbox}
\usepackage{comment}
\usepackage{booktabs}
\usepackage{graphicx}
\usepackage{makecell}
\usepackage{multirow}
\usepackage{subcaption}
\usepackage{tikz}
\usepackage{caption}
\usepackage{flushend}
\usepackage[shortcuts]{extdash}
\usepackage{balance}
\usepackage{paralist}
\usepackage{enumitem}
\usepackage{siunitx}
\usepackage{listings}
\usepackage{xcolor}
\usepackage[capitalize,nameinlink]{cleveref}

\mathchardef\mhyphen="2D 

\newcommand{\sysname}{Murakkab\xspace}
\newcommand{\eg}{\emph{e.g.,}\xspace}
\newcommand{\etc}{etc.\@\xspace}

\newcommand{\ie}{\emph{i.e.,}\xspace}

\newcommand{\myparagraph}[1]{\vspace{\smallskipamount}\noindent\textbf{#1.\xspace}}

\newtoggle{showmarks}
\toggletrue{showmarks}

\iftoggle{showmarks}{
  
  \newcommand{\oldtext}[1]{\leavevmode\textcolor{red}{OLD: #1}}
  
  \newcommand{\inigo}[1]{\textcolor{purple}{IG: #1}}
  \newcommand\esha[1]{{\color{cyan}{\bf EC: #1}}}
  \newcommand\rf[1]{\textcolor{olive}{RF: #1}}
}{
  
  \newcommand{\oldtext}[1]{\unskip}
  
  \newcommand{\inigo}[1]{\unskip}
  \newcommand\esha[1]{\unskip}
  \newcommand\rf[1]{\unskip}
}

\lstdefinestyle{custompython}{
  language=Python,
  basicstyle=\ttfamily\footnotesize,
  keywordstyle=\color{blue}\bfseries,
  commentstyle=\color{green!50!black},
  stringstyle=\color{red!60!black},
  backgroundcolor=\color{gray!10},
  frame=single,
  breaklines=true,
  postbreak=\mbox{\textcolor{red}{$\hookrightarrow$}\space},
  otherkeywords={[2]self, db_type, params, name, resources, key, inputs, system_prompt, user_prompt, description, constraints, workflow},
  morekeywords={[2]self, db_type, params, name, tasks, resources, key, inputs, system_prompt, user_prompt, description, constraints, workflow},
  keywordstyle={[2]\color{magenta}},
  captionpos=b,
  morecomment=[l]{\#},
  deletekeywords={self, db_type, params, name, tasks, resources, key, inputs, system_prompt, user_prompt, description, constraints, workflow}
}

\begin{document}

\title{Towards Resource-Efficient Compound AI Systems}

\author{
  Gohar~Irfan~Chaudhry$^{1}$, 
  Esha~Choukse$^{2}$, 
  Íñigo~Goiri$^{2}$, 
  Rodrigo~Fonseca$^{2}$, 
  Adam~Belay$^{1}$,
  Ricardo~Bianchini$^{2}$
}

\affiliation{%
  $^{1}$MIT CSAIL \hspace{1cm} $^{2}$Microsoft Azure
}

\begin{abstract}
Compound AI Systems, integrating multiple interacting components like models, retrievers, and external tools, have emerged as essential for addressing complex AI tasks.
However, current implementations suffer from inefficient resource utilization due to tight coupling between application logic and execution details, a disconnect between orchestration and resource management layers, and the perceived exclusiveness between efficiency and quality.

We propose a vision for resource-efficient Compound AI Systems through a \emph{declarative workflow programming model} and an \emph{adaptive runtime system} for dynamic scheduling and resource-aware decision-making.
Decoupling application logic from low-level details exposes levers for the runtime to flexibly configure the execution environment and resources, without compromising on quality.
Enabling collaboration between the workflow orchestration and cluster manager enables higher efficiency through better scheduling and resource management.

We are building a prototype system, called \textbf{\textit{\sysname{}}}, to realize this vision.
Our preliminary evaluation demonstrates speedups up to $\sim 3.4\times$ in workflow completion times while delivering $\sim 4.5\times$ higher energy efficiency, showing promise in optimizing resources and advancing AI system design.
\end{abstract}

\renewcommand\footnotetextcopyrightpermission[1]{}
\settopmatter{printacmref=false}
\maketitle
\pagestyle{plain}

\section{Introduction}

\begin{figure}[t]
    \includegraphics[width=\columnwidth]{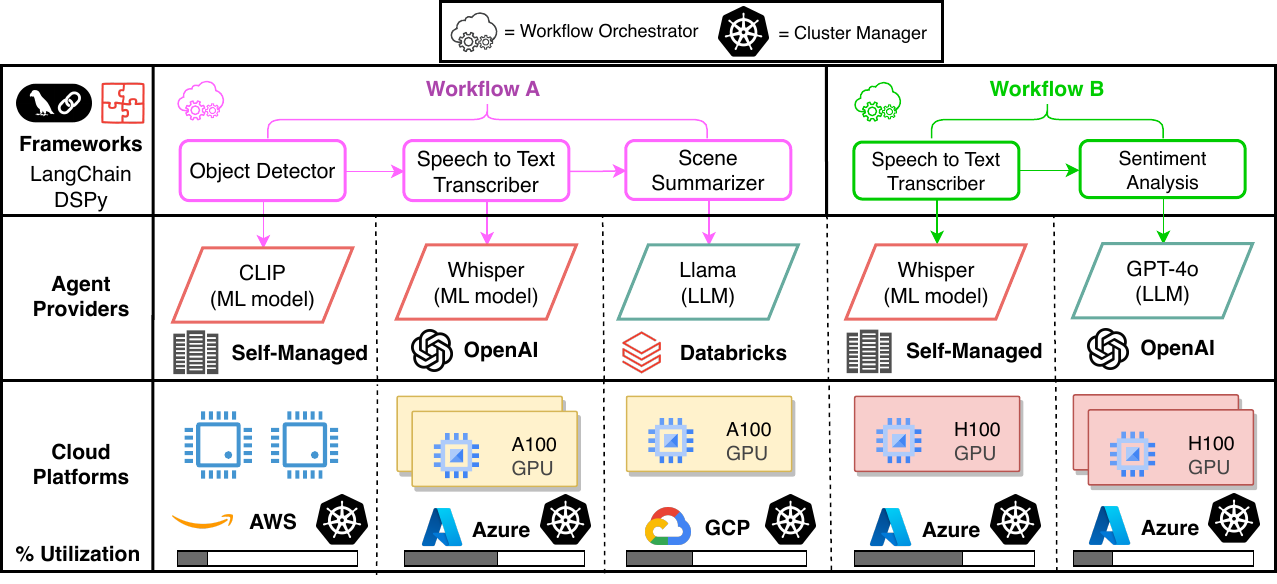}
    \caption{\footnotesize Today programmers use frameworks to call agents from \emph{different} providers hosted on \emph{multiple} cloud platforms. The rigid coupling between all layers of the system results in inefficiencies.}
    \label{fig:entities}
    \vspace{-15pt}
\end{figure}
 
\begin{figure}[t]
    \includegraphics[width=\columnwidth]{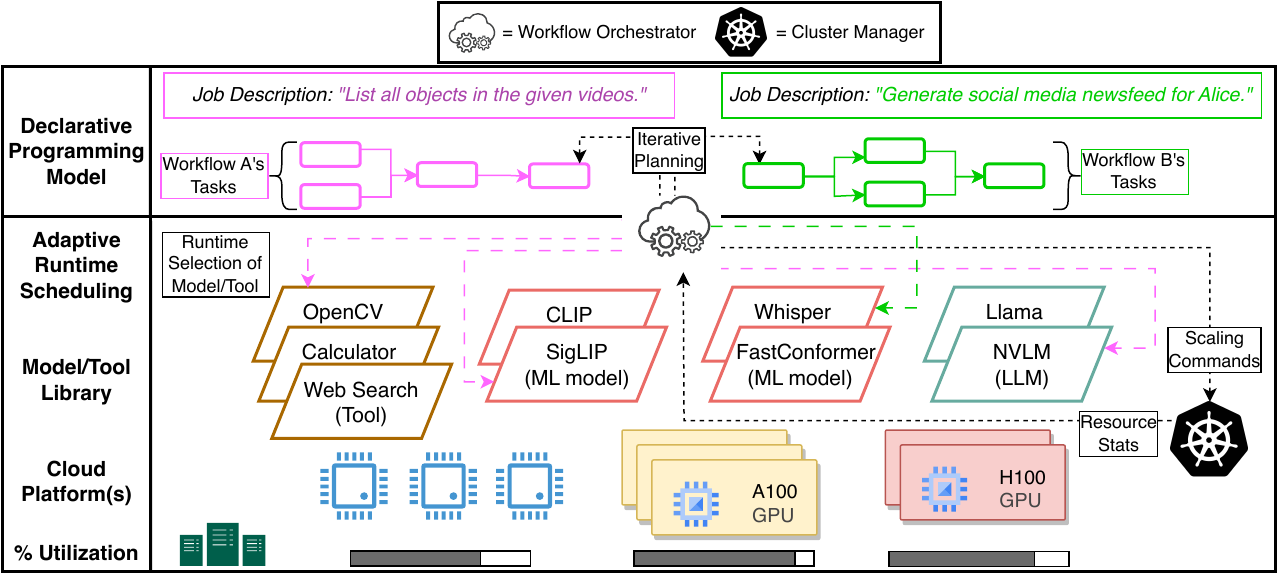}
    \caption{\footnotesize We envision \emph{fungible workflows} with high-level descriptions, managed jointly by the Workflow Orchestrator and Cluster Manager. This allows higher resource multiplexing between independent workflows to improve efficiency.}
    \label{fig:vision}
     \vspace{-10pt}
\end{figure}

A Compound AI System is ``a system that tackles complex tasks using multiple interacting components, including multiple calls to different AI models, retrievers, or external tools''~\cite{compound-ai-blog}.
With increasing task complexity and model capabilities, the workflows promise to grow deeper, more complex, and self-morphing with self-improving agents.
Today, workflows are designed by explicitly defining the components, their interactions, and the allocation of resources.

While effective for many tasks, these workflows in practice frequently suffer from inefficient resource utilization.
\Cref{fig:entities} shows a typical Compound AI workflow deployment, with multiple stages. Each stage in the workflow comprises of three key entities:

\vspace{\smallskipamount}\noindent\emph{Programming Frameworks}
    to create workflows by composing \emph{agents} including LLMs, ML models, and tools.
    They have \emph{workflow orchestrators} that may decide agent execution order, optimize prompts for LLMs, process intermediate outputs and provide memory to stateless models \etc{}
    Examples include LangChain~\cite{langchain}, LlamaIndex~\cite{llamaindex} and DSPy~\cite{dspy}.

\vspace{\smallskipamount}\noindent\emph{Agent Providers}
    offer specific models, tools, or vector databases \etc{} typically through REST APIs, to invoke from workflows.
    These could be proprietary (\eg{} GPT 4o~\cite{gpt4o}) or open-source (\eg{} Llama~\cite{llama3}) models.
    They may also provide additional features like analytics, data storage and management \etc{}
    Example providers include OpenAI~\cite{openai_llms} and Databricks~\cite{databricks_llm_serving}.

\vspace{\smallskipamount}\noindent\emph{Cloud Platforms}
    rent out hardware infrastructure like GPUs, CPUs, and storage for running models, tools and vector databases \etc{}
    They may individually run \emph{cluster managers} to monitor reasource utilization, take scaling decisions and perform load balancing among instances \etc{}
    Example platforms include AWS~\cite{aws}, Azure~\cite{azure}, and GCP~\cite{gcp}.

Entities in the stack have differing efficiency objectives:
(a) programmers prioritize result quality,
(b) agent providers aim to offer diverse tools at lower costs, and
(c) cloud platforms focus on maximizing utilization and profitability.
Workflows spanning multiple providers and platforms add complexity.

\myparagraph{Challenges} Given the state-of-the-art AI workflows, key challenges include:
\begin{enumerate}[leftmargin=*]
\item Tight coupling of application logic with execution configurations (\eg{} model and hardware) restricts efficient alternatives.
\item Disconnects between workflow orchestration and cluster management (often separately owned) result in suboptimal scheduling.
\item Balancing resource efficiency (\eg{} cost, power) with end-to-end result quality (model accuracy) is difficult, as over-provisioning fragments resources and under-provisioning degrades performance.
\end{enumerate}

These inefficiencies affect all entities, as workflow users and agent providers either pay for unused resources or experience degraded performance, while cloud platforms suffer from poor resource utilization.

\myparagraph{Our Work}
We believe that future Compound AI Systems will become increasingly complex.
These systems are likely to integrate self-improving agents, dynamic workflows for open-ended tasks, unpredictable execution flows, and an ever-expanding library of models and tools.
Managing such systems efficiently requires rethinking the entire stack and redefining the roles of its 
layers.

We present \sysname{}, a system that enhances resource efficiency through \emph{fungible workflows} and dynamic scheduling (\Cref{fig:vision}). Key components include:
(a) a \emph{declarative workflow programming model} that abstracts model, tool, and hardware choices, simplifying development and enabling flexibility, and
(b) an \emph{adaptive runtime system} that integrates workflow orchestration and cluster management for resource-aware scheduling and proactive resource management.
Preliminary evaluation of \sysname{} shows speedups of $\sim 3.4\times$ in workflow completion times with $\sim 4.5\times$ higher energy efficiency.

\begin{figure}[t]
\end{figure}

\section{Today's Imperative Workflows}
Workflows in Compound AI Systems today are typically expressed through imperative programs that contain:
(1) \emph{the system flow} specifying the components and their interaction,
(2) \emph{model types and configuration details} to implement each component and any model/tool specific parameters,
(3) \emph{resources for each component} in terms of hardware configuration, and
(4) \emph{pricing tiers} in terms of performance guarantees (\eg{} token generation throuhgput \etc{})

\begin{figure*}[t]
\lstset{style=custompython}
\begin{lstlisting}[
language=Python,
caption={Video Understanding workflow defined for today's Compound AI Systems. This requires explicit selection of models/tools and details of the providers offering them (\ie{} API keys). For clarity, we show the resource configuration for each agent in-line, although these details are typically specified when signing up with agent providers and/or cloud platforms.},
label={lst:workflow_prog_today}
]
# Define the components (models/tools), hyperparameters, resource specifications and pricing tiers
frame_ext = Tool(name="OpenCV", params={sampling_rate: 15}, key=ON_PREM_SSH_KEY, resources={CPUs: 1})
stt       = MLModel(name="Whisper", key=OPENAI_API_KEY, resources={PTUs: 1})
obj_det   = MLModel(name="CLIP", key=AWS_SSH_KEY, resources={CPUs: 2})
summarize = LLM(name="llama", key=DATABRICKS_API_KEY, params={context_len: 4096},
                resources={GPUs: 1, GPU_Type: H100},
                system_prompt="You are an agent that can describe images in detail.",
                user_prompt="Summarize the scenes using frames, detected objects and transcripts.")
# Inputs
videos = ["cats.mov", "formula_1.mov"]
# Describe the data flow between components
result = Workflow(frame_ext -> stt -> obj_det -> summarize).execute()
\end{lstlisting}
\end{figure*}

\Cref{lst:workflow_prog_today} shows such an example of a Video Understanding workflow, based on OmAgent~\cite{zhang2024omagent}.
It defines components (lines 2 to 8) to perform various tasks and their execution flow (line 12).
For example, it has a frame extraction (\texttt{frame\_ext}) component that uses OpenCV~\cite{opencv_library} and has task-specific parameters like sampling rate.
For audio processing, it has a speech-to-text transcription agent (\texttt{stt}) implemented using Whisper~\cite{whisper}.
It uses an LLM, in this case Llama~\cite{llama3}, for summarizing scenes and specifies a context length.
For each of these components, there is either a hardware configuration (\eg{} 1 NVIDIA H100~\cite{nvidia_h100} GPU) or pricing tier (\eg{} 4 Provisioned Throughput Units or PTUs~\cite{azure_provisioned_throughput}).

This workflow tightly integrates the application logic (\eg{} "List objects shown/mentioned in the videos"), the specific models to use (\eg{} Llama), and the resources to allocate (\eg{} GPUs: 1 or PTUs: 4).
In addition to developing application logic, the developer has added burden to configure many hyperparameters and resource specifications.
Often, these selections are suboptimal---there could either be resource stranding or underprovisionig leading to suboptimal performance.
As a result we end up in a situation similar to \Cref{fig:entities}, with rigid cross-layer coupling, that makes it challenging to improve efficiency of such systems.
\section{Efficiency Through Fungibility}

\begin{figure}[t]
\lstset{style=custompython}
\begin{lstlisting}[
language=Python,
caption={Video Understanding workflow defined for \sysname{}. This only requires declaring a high-level job description and optional hints/constraints.},
label={lst:workflow_prog_dynamic}
]
# Define the job in natural language
desc ="List objects shown/mentioned in the videos"
# Optional: Specify sub-tasks in the job
t1 = "Extract frames from each video"
t2 = "Run speech-to-text on all scenes"
t3 = "Detect objects in the frames"
# Inputs
videos = ["cats.mov", "formula_1.mov"]
# Execute
result = Job(description=desc, inputs=videos,
             tasks=[t1, t2, t3],
             constraints=MIN_COST).execute()
\end{lstlisting}
\vspace{-15pt}
\end{figure}

\Cref{fig:vision} shows our vision for Compound AI Systems, where application logic is decoupled from execution details.
Developers focus on application logic, without managing model selection, updates, or resource demands.
The runtime dynamically generates task graphs from high-level descriptions, mapping tasks to models and tools for efficient resource multiplexing while maintaining quality.
It leverages application fungibility to optimize for efficiency at runtime~\cite{ruan2023quicksand}. 

We draw inspiration from the evolution of SQL and the query optimizations that are enabled by its declarative nature~\cite{jarke1984optimization, chaudhuri1998optimization}.
Recent work has taken initial steps in exploring this for AI systems~\cite{liu2024declarative, anderson2024designllmpoweredunstructuredanalytics, madden2024querying}, albeit for narrow use cases and with limited focus on resource efficiency and multi-tenancy.
We aim to bridge this gap with \sysname{}.

\subsection{Declarative Workflow Programming}
\sysname{} promotes high-level declarative workflow specifications for two reasons:
(1) it frees developers from managing low-level implementation details, and
(2) it enhances workflow flexibility by allowing dynamic selection of models, tools, and resources at runtime to improve efficiency.

\Cref{lst:workflow_prog_dynamic} shows the same Video Understanding workflow defined for \sysname{}.
The programmer provides a natural language description of the job (\texttt{desc}) and the required inputs for the job (\texttt{inputs}).
The programmer may optionally assist the system by specifying sub-tasks (lines 4 to 6) that need to be performed to accomplish the job.
However, if these tasks are not provided or are insufficient, an orchestrator LLM decomposes the job into smaller tasks based on the provided job description.
It also identifies the relationship between tasks and generates the corresponding internal representation as a directed acyclic graph (DAG) where the nodes represent agents, and edges represent dataflow between them.
Moreover, the programmer can also specify high-level constraints for performance or quality (\eg{} \texttt{MIN\_COST} would let the system decide an execution strategy that minimizes execution cost of the workflow, potentially in exchange for latency.)
In the future, we plan to support multiple constraints with a priority ordering.

Note that specific models, tools, and hardware resources are abstracted from the developer while allowing \sysname{} to dynamically select them at runtime.

\subsection{Adaptive Runtime Scheduling}

\begin{table*}[t]
    \renewcommand{\arraystretch}{0.85}
    \centering
    \footnotesize
    \begin{tabular}{@{}c|c|c|ccc|c@{}}
    \toprule
    \textbf{Parameter} & \textbf{Category} & \textbf{Selection} & \textbf{\$ Cost} & \textbf{Power} & \textbf{Latency} & \textbf{Quality} \\ \midrule
    GPU Generation & Hardware Type & Newer & Higher & Higher & Lower/No Change & No Change \\
    CPU vs GPU & Hardware Type & CPU & Lower & Lower & Lower & No Change \\
    Task Parallelism & Resource Amount & More Fan Out & Higher & Higher & Lower & No Change \\
    Execution Paths & Resource Amount & More Paths & Higher & Higher & Higher/No Change & Higher/No Change \\
    Model/Tool & Agent Implementation & More Parameters & Higher & Higher & Higher & Higher/No Change \\ \bottomrule
    \end{tabular}
    \caption{Optimization parameters and their impact on efficiency and quality.
    For simplicity we show a single selection.
    }
    \label{tab:parameters}
    \vspace{-10pt}
\end{table*}

\Cref{tab:parameters} outlines the parameters \sysname{} adjusts to optimize monetary cost, power consumption, and execution latency while measuring their impact on workflow result quality.
Below, we describe \sysname{}'s adaptive workflow execution and how these parameters guide scheduling decisions.

\myparagraph{Job Decomposition}
Using a declarative programming model requires \emph{lowering} the high-level job specification into actionable tasks.
For dynamic workflows, \sysname{} uses LLMs to decompose a job description into a set of tasks, following the ReAct~\cite{yao2023reactsynergizingreasoningacting} approach.
The LLM orchestrates the execution order of tasks in the workflow and outputs a DAG.

\myparagraph{Task-to-Agent Mapping}
\sysname{} maintains a flexible library of agents, detailing their names, functionalities, and schemas (\eg{} function arguments).
The orchestrator uses this library and task descriptions to map tasks to suitable agents.
For example, with NVLM~\cite{nvlm2024} as the orchestrator LLM, \sysname{} provides the agent library via the \textit{system prompt}~\cite{openai_api_reference} and task descriptions via the \textit{user prompt}.
This enables the LLM to assign agents to tasks.
\sysname{} then supplies task metadata and input details to the LLM, requesting a \textit{tool call}~\cite{openai_function_calling_guide} for the selected agent.
The LLM generates an executable code snippet with the necessary arguments to invoke the agent directly.
For example, given the task \texttt{"Extract frames from each video"} and appropriate metadata, the LLM may generate the following tool call:
\texttt{FrameExtractor(start\_time=0, end\_time=60s, num\_frames=10, file="cats.mov")}.

\myparagraph{Model/Tool Selection}
The library may contain multiple models or tools that support a given agent interface.
For instance, the \texttt{Speech-to-Text} agent can be implemented using Whisper, DeepSpeech, Fast Conformer~\cite{rekesh2023fastconformerlinearlyscalable}, and others.
Each differs in response quality, performance and resource requirements.
\sysname{} generates an \emph{execution profile} for each model/tool and hardware resource pair when a new one is added to the library---the profile captures an efficiency vs quality tradeoff.
Efficiency metrics include cost, power consumption, and latency.
At runtime, \sysname{} selects the model/tool and resources that maximize efficiency while meeting the target quality.

\myparagraph{Resource Allocation}
Cloud platforms provide various hardware SKUs, including different GPU and CPU types with dynamic availability (\eg{} Spot VMs~\cite{azure_spot_vms}, Harvest VMs~\cite{ambati2020providing}).
Models and tools can run on a range of these hardware types.
For instance, some models perform better on newer GPUs (\eg{} NVIDIA H100~\cite{nvidia_h100}) with higher FLOPS, while others see no significant benefit.
Similarly, certain models run efficiently on CPUs, whereas others may be too slow to execute practically.
\sysname{} dynamically allocates resources to the models and tools by using their execution profiles and real-time resource availability metrics from the Cluster Manager.

\myparagraph{Execution Paths}
The Orchestrator leverages parallelism within the workflow, assigning additional resources to agents for improved performance by breaking tasks into sub-tasks and executing them in parallel.
For example, \texttt{FrameExtractor} can split a video into smaller chunks for parallel extraction when resources are available.
In some cases, the Orchestrator may explore multiple execution paths in parallel to enhance result quality.
For example, in Chain-of-Thought~\cite{wei2023chainofthoughtpromptingelicitsreasoning} workflows, allocating more resources allows exploration of additional reasoning paths, with the final result determined by top-k outputs.
These decisions are based on cost constraints and real-time resource availability.

\myparagraph{Workflow-Aware Cluster Management}
Typically used cluster management systems (\eg{} Kubernetes~\cite{kubernetes} for microservices) are not well-suited for Compound AI Systems as they lack the necessary insights for workflow scheduling and orchestration, which are integral to such systems (\Cref{fig:entities}).
\sysname{} bridges this gap by facilitating information exchange between workflow scheduling and cluster management (\Cref{fig:vision}).
It exposes workflow DAGs to the Cluster Manager, providing visibility into completed and upcoming tasks. This enables the Cluster Manager to rebalance resources across models and tools more effectively.
With this enhanced visibility, the Cluster Manager can make better scaling and resource allocation decisions.
For example, if no workflows are expected to require a \texttt{Speech-To-Text} agent soon, it can reallocate GPU resources from Whisper to Llama in anticipation of increased demand.

We wish to incorporate learning from prior cluster management research~\cite{mage, quasar, paragon} to efficiently use heterogeneous hardware, offer QoS and perform online scheduling.

\myparagraph{Resource-Aware Worfklow Orchestration}
The Workflow Orchestrator continuously receives stats from the Cluster Manager including idle resources, per-model or tool resource consumption and any harvestable resources like Spot Instances~\cite{azure_spot_vms,ambati2020providing}.
The Orchestrator prefers selecting models/tools that are already running or for which there are enough resources available to handle incoming requests.
Resource efficiency is improved by maximizing resource multiplexing and minimizing fragmentation.

Thus, integrating the Workflow Orchestrator and Cluster Manager is crucial to unlocking efficient Compound AI Systems.
\sysname leverages this design to the fullest.

\subsection{\sysname{} Overheads}
\sysname{} has several overheads associated with it.
(a) \emph{Profiling}: To be able to offer different resource configurations, we need to profile the agents and tools on different hardware and configurations. However, this profiling is amortized over the lifetime of all the workflows that use a particular agent or tool.
(b) \emph{DAG Creation}: Task understansing from a natural language prompt, and DAG creation requires LLM queries. However, these are short input and short output queries, that take less than 1\% of the execution time of the target AI workflows.
(c) \emph{Configuration Search}: The search space across the levers mentioned in \cref{tab:parameters} can easily explode. Therefore, we are working on strategies to prune the space with greedy search using hierarchy of optimization functions.
\section{Evaluation}
Our evaluation examines whether \sysname{} can take advantage of the fungibility of the declarative workflow to identify different execution configurations using the levers in \cref{tab:parameters} and make a selection based on their efficiency/performance trade-off.
We run the Video Understanding workflow derived from OmAgent~\cite{zhang2024omagent} as shown in \Cref{lst:workflow_prog_today} as the baseline and \Cref{lst:workflow_prog_dynamic} on \sysname{}.
The execution output and accuracy are the same in all comparisons.

\myparagraph{Setup}
We run our experiments on two Azure VMs (Standard\_ND96amsr\_A100\_v4~\cite{azure_ndma100v4}) each with 96 AMD EPYC 7V12 vCPUs and 8 NVIDIA A100 (80GB) GPUs~\cite{nvidia_a100}.
We use an OpenCV~\cite{opencv_library}-based frame extractor (CPUs), NVLM~\cite{nvlm2024} as the LLM for frame summarization (8 GPUs for text completion and 2 GPUs for embeddings to insert in a VectorDB for question/answering), CLIP~\cite{clip} for Object Detection (CPUs) and Whisper~\cite{whisper} for Speech-to-Text transcription (1 GPU).

\begin{figure}[t]
    \centering
    \includegraphics[width=.98\columnwidth]{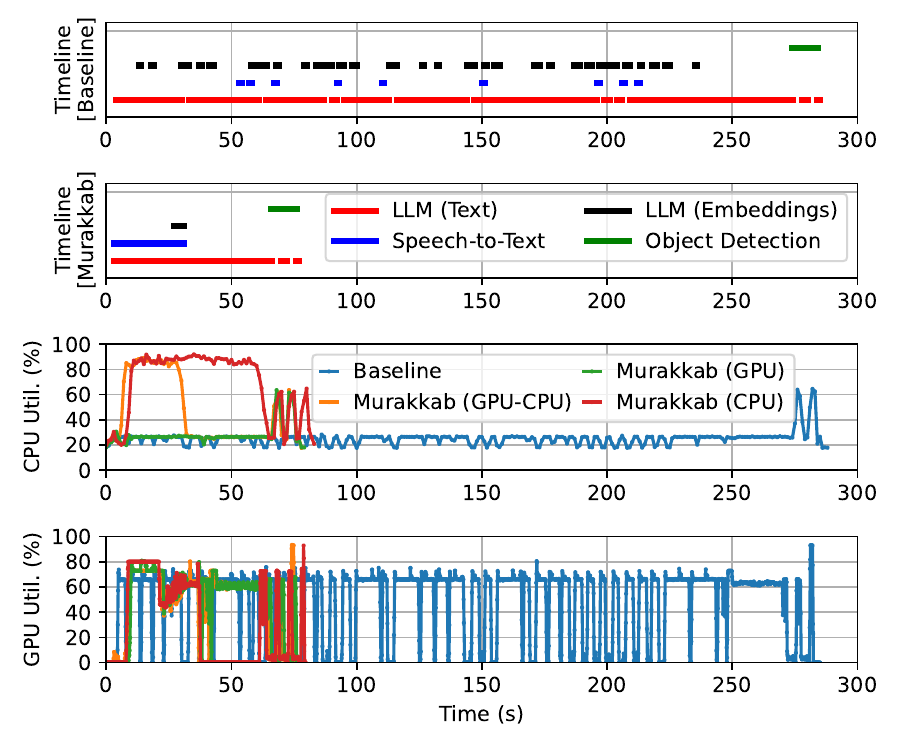}
    \caption{Execution traces of the Video Understanding workflow. \sysname{} can adjust between multiple configurations and deliver a $\sim 3.4\times$ speedup with higher resource efficiency.}
    \label{fig:eval}
\end{figure}

\myparagraph{Baseline}
Derived from OmAgent~\cite{zhang2024omagent}, the baseline workflow specifies a fixed execution without any intra-task parallelism or opportunity to utilize idle resources.
Each scene and its constituent frames are processed sequentially.
\Cref{fig:eval} shows the results---this workflow completes in 283$s$ and severely underutilizes resources.

\myparagraph{\sysname{}}
We use NVLM as the orchestrator, and generate a DAG for the workflow to identify parallelism opportunities.
Speech-to-Text (STT) is identified as the main dependency for the later stages.
Based on STT's internal tasks and DAG, \sysname{} is able to perform three optimizations. It:
(a) executes STT transcription for multiple scenes in parallel (leveraging dataflow structure from the DAG),
(b) parallelizes intra-scene frame summarization (leveraging underutilized GPUs), and
(c) adjusts the resource configuration for STT (using execution profiles for Whisper).
We show execution traces from the various resource configurations that \sysname{} can choose for STT---1 GPU (similar to the baseline), 64 CPU cores, and a combination of GPU and CPUs.
\Cref{fig:eval} shows the results---\sysname{} can complete the workflow between 77--83$s$, a $\sim 3.4\times$ speedup over the baseline. 

\Cref{tab:energy_consumption_time} shows the energy consumption and execution times of each configuration.
For simplicity we only measure the GPU energy consumption since that is the dominant source in the system (rated $16\times$ higher than the CPU power).
Executing STT entirely on CPUs results in the lowest energy consumption (34$Wh$), while executing it entirely on GPUs results in the fastest execution (77$s$).
\sysname{} selects the CPU configuration to satisfy the \texttt{MIN\_COST} constraint (\Cref{lst:workflow_prog_dynamic}), resulting in $\sim 4.5\times$ higher energy efficiency. 

\begin{table}[t!]
    \renewcommand{\arraystretch}{0.65}
    \centering
    \begin{tabular}{c|c|c}
    \toprule
    \textbf{Speech-to-Text Config.} & \textbf{Energy (Wh)} & \textbf{Time (s)} \\
    \midrule
    Baseline                          & 155                               & 285               \\
    \sysname{} CPU                    & 34                                & 83                \\
    \sysname{} GPU                    & 43                                & 77                \\
    \sysname{} GPU + CPU              & 42                                & 77                \\
    \bottomrule
    \end{tabular}
    \caption{Energy and execution time of each configuration.}
    \label{tab:energy_consumption_time}
    \vspace{-15pt}
\end{table}

\section{Discussion}

\myparagraph{AI Workflows-as-a-Service (AIWaaS)}
We envision a future for Compound AI Systems where developers focus solely on application logic, without needing to manage model or resource details.
Similar to Functions-as-a-Service (FaaS)~\cite{jonas2019cloud}, where the runtime system handles resource allocation and load balancing, we propose an AI Workflows-as-a-Service (AIWaaS) model with similar capabilities.
This can improve efficiency, lower operational costs, and make AI systems more accessible and easier to maintain.
Applications will not need rewriting (\eg{} prompt engineering, workflow recreation) when new models or tools are available---the runtime system will transparently adopt newer implementations and resources as needed.

\myparagraph{Quantifying and Controlling Quality}
Cost-quality tradeoffs are well-studied for single-model queries (e.g., FrugalGPT~\cite{chen2023frugalgptuselargelanguage}), but end-to-end workflows pose unique challenges. Model interactions cause cascading effects, making it costly and impractical to evaluate all combinations. We explore methods to narrow the search space by identifying stages with the greatest impact on cost and accuracy. Additionally, \emph{hallucinations} in early stages can derail workflows, highlighting the need for more correctness checkpoints and tools for quality control.

\myparagraph{Proprietary Models and Agents}
Integrating agent providers and cloud platforms into a unified entity can improve resource efficiency in Compound AI Systems.
However, proprietary models often cannot be exported, requiring external API calls to third-party models.
This may reduce resource efficiency due to limited visibility into third-party resource usage.
Further research is needed to determine when offloading tasks to third-party providers is more beneficial than using local models/tools, albeit with lower quality.

\myparagraph{Multi-cloud Compound AI Systems}
Greater control over hardware resources is easier when the cloud platform and workflow execution service are managed by the same entity (\eg{} Azure ML~\cite{azure_machine_learning}, AWS SageMaker~\cite{aws_sagemaker}).
However, using multiple cloud platforms~\cite{stoica2021skycomputing} can reduce costs and offer a wider variety of hardware (\eg{} Google TPUs~\cite{jouppi2023tpuv4opticallyreconfigurable}).
This is possible if each platform exposes resource utilization metrics, allowing systems like \sysname{} to manage resources across clouds and schedule tasks efficiently.

\section{Conclusion}

This work highlights inefficiencies in existing Compound AI Systems and identifies key challenges limiting resource optimization.
To overcome these, we propose a reimagined architecture featuring fungible workflows, dynamic scheduling, and adaptive resource management.
By unifying workflow orchestration with cluster management, our system enhances resource utilization, reduces operational costs, and maintains or improves result quality.
Our preliminary evaluations show significant gains in efficiency, validating our approach.
Looking ahead, our AIWaaS vision aims to simplify AI application development and make AI systems more accessible and sustainable across diverse use cases.

\balance
\bibliographystyle{ACM-Reference-Format}
\bibliography{paper}

\end{document}